\documentclass[prl,twocolumn,superscriptaddress,longbibliography,notitlepage]{revtex4-1}
\usepackage{amssymb}
\usepackage{amsmath}
\usepackage{graphicx}
\usepackage{hyperref}

\begin{document}

\title{The Quantum Cocktail Party Problem}

\author{Xiao Liang}
\affiliation{Laboratory of Quantum Information, University of Science and Technology of China, Hefei, 230026, China}
\affiliation{Institute for Advanced Study, Tsinghua University, Beijing, 100084, China}

\author{Yadong Wu}
\affiliation{Institute for Advanced Study, Tsinghua University, Beijing, 100084, China}

\author{Hui Zhai}
\email{hzhai@tsinghua.edu.cn}
\affiliation{Institute for Advanced Study, Tsinghua University, Beijing, 100084, China}

\begin{abstract}
The cocktail party problem refers to the famous selective attention problem of how to find out the signal of each individual sources from signals of a number of detectors. In the classical cocktail party problem, the signal of each source is a sequence of data such as the voice from a speaker, and each detector detects signal as a linear combination of all sources. This problem can be solved by a unsupervised machine learning algorithm known as the independent component analysis. In this work we propose a quantum analog of the cocktail party problem. Here each source is a density matrix of a pure state and each detector detects a density matrix as a linear combination of all pure state density matrix. The quantum cocktail party problem is to recover the pure state density matrix from a number observed mixed state density matrices. We propose the physical realization of this problem, and how to solve this problem through either classical Newton's optimization method or by mapping the problem to the ground state of an Ising type of spin Hamiltonian. 

\end{abstract}

\maketitle

\textbf{Introducation.}

The cocktail party problem refers to the phenomenon that the brain of a listener can focus on a single voice while filtering out a range of other voices in a multi-talker situation, say, in a cocktail party \cite{Review}. This selective attention problem is first defined as the ``cocktail party problem" (CPP) by C. Cherry in 1953 \cite{Cherry}. For several decades, it is an important research subject for both neuroscience to understand how human or animals solve this problem and computer science to design algorithms to solve this problem. During recent years, machine-learning based approach to solve the CPP is essential for many industrial applications such as automated speech recognization. The independent component analysis (ICA) is such an algorithm particularly suitable for the CPP. The CPP can also found its application in physical science such as astrophysics data analysis \cite{CPP_physics1,CPP_physics2}. Recently, it has also been proposed to use the spirt of CPP and ICA method to extract the eigen frequency of a quantum system from a dynamical probe \cite{Wu}.

\begin{figure}[t]
\begin{center}
\includegraphics[width=0.9\columnwidth]{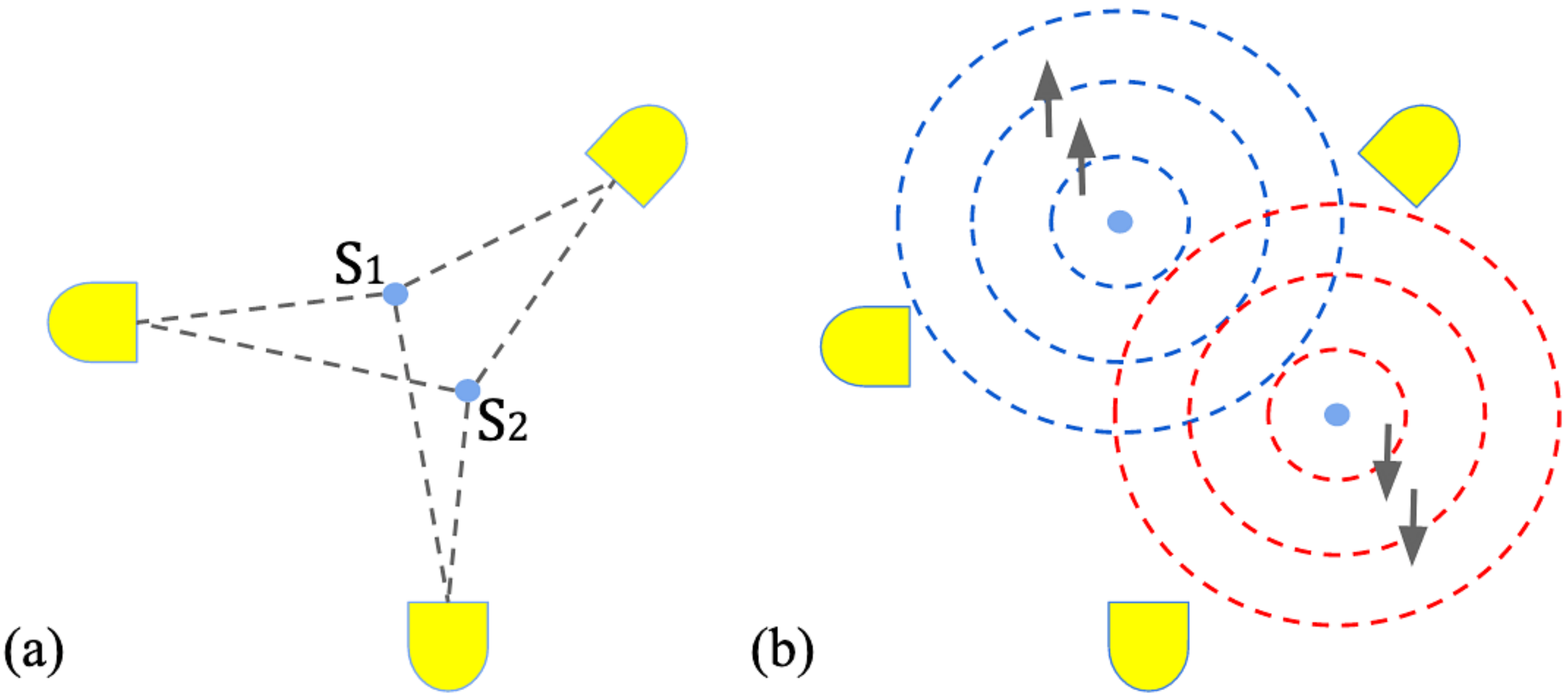}
\caption{Schematic of the cocktail party problem. (a) Classical case. The voices emitted from different source are mixed and detected by detectors. (b) Quantum case. The wave functions from different sources mix spatially and the density matrix at different places are detected by detectors.  }
\label{Fig:CPP_illustrate}
\end{center}
\end{figure}

Let us first briefly review the classical CPP (c-CPP). Considering $N$-independent speakers in a room, they speak simultaneously and the voice of each speaker is a source denoted by a sequence $s_i(t)$ ($i=1,\dots,N$). There are also $M$ detectors in the room. Each detector detects a signal $x_j(t)$ ($j=1,\dots, M$) that is considered to be a linear combination of all sources $s_i(t)$, as schematically shown in Fig. \ref{Fig:CPP_illustrate}(a). That is to say, we have a $M\times N$-dimensional matrix $A$ and 
\begin{equation}
x_j(t)=\sum\limits_{ji}A_{ji}s_i(t).
\end{equation}
To concentrate on one of the speakers, it means that we should find out $A^{-1}$ such that we can determine $s_i(t)$ as $s_i(t)=\sum_j(A^{-1})_{ij}x_j(t)$ from the signals of all detectors. For human, we only have two ears which means the number of detectors is two. But for computer algorithm problem, we make the situation simpler by considering that there are more detectors than sources, that is, $M>N$. Even though, this is still an ill-defined problem if no information of the source is known. In practices, we utilize the information that being voice of an individual speaker, each source $s_i(t)$ displays certain feature and is more regular than a mix of several voices. By performing statistics over $t$ for each sequence $s_i(t)$, we can determine the entropy of the sequence and we use the criterion that the entropy of each sequence should be minimized to determine each $s_i(t)$. This is how we solve the CPP with the ICA method \cite{ICA}.

In this work we will propose a quantum analogy of the CPP, termed as the quantum cocktail party problem (q-CPP). We will discuss how to solve the q-CPP with an analogy of the ICA method. We should also present a mathematical statement that can help us to map the loss function to a Hamiltonian of Ising spins. Though by classical Monte Carlo, we show that the ground state spin configuration can solve the q-CPP, we point out that this spin Hamiltonian can be solved more efficiently by quantum methods, for example, by quantum simulation and quantum annealing. 

\textbf{Results.}

\textit{Quantum CPP.} Here we first propose the q-CPP. We consider $N$ different sources, and each source $s_i$ to be a density matrix of a pure state as $|\phi_i\rangle\langle\phi_i|$, where $|\phi_i\rangle$ is a normalized quantum state in a Hilbert space with dimension $d$. There are $M$ number of detectors, and the signal $x_j$ detected by each detector is a density matrix of a mixed state denoted by $\rho_j$ as 
\begin{equation}
\rho_j=\sum\limits_{ji}A_{ji}|\phi_i\rangle\langle\phi_i|. \label{q_CPP_dm}
\end{equation}  
We also normalize $\rho_j$ to be trace unity, which require $\sum_{i}A_{ji}=1$ for all $j$. The q-CPP is defined as that, suppose that we know sufficient number of $\rho_j$, whether one can find out $A_{ji}$ to recover each $|\phi_i\rangle\langle\phi_i|$. 

Here we will briefly discuss the uniqueness of the solution. First of all, we should emphasize that in order to ensure the solution is unique, it is important to require that different $|\phi_i\rangle$ are \textit{not} orthogonal to each other. Secondly, when the number of detector increases by one, the constraints increases by $d^2$ and the free parameters increases by $N-1$, so we will consider the situation that $d^2>N$. Lastly, it is always good to have sufficient number of detectors, normally we consider $M>N$. We do not rigorously prove the uniqueness of the solution, but we find that in practices, generically we always find unique solution when these conditions are satisfied.  

A physical realization of the q-CPP can be proposed as follows. Let us consider a particle whose internal Hilbert space is a product of two degrees of freedom as $\mathcal{H}=\mathcal{H}_A\times \mathcal{H}_B$. The dimensionality of $\mathcal{H}_A$ is $d$ and the dimensionality of $\mathcal{H}_B$ is $N$, and $\{|s_i\rangle\}$ ($i=1,\dots,N$) forms a complete set of bases in $\mathcal{H}_B$. A wave function $|\Phi\rangle$ can be generally expanded in these bases as
\begin{equation}
|\Phi\rangle=\sum\limits_{i=1}^{N}\varphi_i({\bf r})|\phi_i\rangle|s_i\rangle.  \label{wf}
\end{equation}
For a more physical picture, one can consider Eq. \ref{wf} as particle emitted from $N$ different sources, and the wave function in $\mathcal{H}_B$ is $|s_i\rangle$ for particle emitted from the source-$i$, as schematically shown in Fig. \ref{Fig:CPP_illustrate}(b). If we place $M$ detectors in different places ${\bf r}_i$ and the quantum measurement traces out the Hilbert space $\mathcal{H}_B$, it results in $M$ different density matrix as Eq. \ref{q_CPP_dm} with $A_{ji}=|\varphi_i({\bf r}_j)|^2$, thus, it is also natural to require $A_{ji}$ to be positive numbers. In practices, these density matrices can be constructed by the quantum state tomography. We consider the situation that both the wave function $\varphi({\bf r})$ and $|\phi_i\rangle$ are unknown. The q-CPP is to determine them from $\rho_j$ ($j=1,\dots,M$).

To find out the pure state, the most important information we use here is that the density matrix of a pure state has the property that $\rho^2=\rho$. Thus, the scheme is to find out a proper combination of $\rho_j$ as $\rho=\sum_{j=1}^{M}w_j\rho_j$ with the normalization condition $\sum_j w_j=1$ that can minimize $|\rho^2-\rho|$. This is equivalent to say, we define the loss function as
\begin{align}
&\mathcal{F}=\sum\limits_{mn}|(\rho^2-\rho)_{mn}|^2\nonumber\\
&=\sum\limits_{mn}|\sum\limits_{j=1}^{M} (\rho^2_j w^2_j-\rho_j w_j)_{mn}+\sum\limits_{i\neq j=1}^{M}(\rho_i\rho_j)_{mn}w_i w_j|^2.
\label{loss}
\end{align}
A comparison between c-CPP and q-CPP is summarized in the Table \ref{Tab:comparison}.

\begin{table}
\centering
\begin{tabular}{|c|c|c|}
\hline
                                      & $\textit{Classical CPP}$                                 & $\textit{Quantum CPP}$              \\ \hline
$\textit{Source}$        & $\text{Voice of each speaker}$ & $\text{Each pure state} $        \\ \hline
$\textit{Detector}$      & $\text{Mixed voices}$          & $\text{Mixed state}$              \\ \hline
$\textit{Loss function}$ & $\text{Minimizing entropy}$    & $\text{Minimizing }$$|\rho^2-\rho|$ \\ \hline
\end{tabular}
	\caption{A comparison between the classical and the quantum cocktail party problem in term of different definition of source, different role of detector and different loss functions.   }
	\label{Tab:comparison} 
\end{table}

\textit{Optimization with Newton's Method.} We firstly use the classical Newton's method to optimize the loss function $\mathcal{F}$, and the update rule of $\textbf{w}$ is:
\begin{equation}
\bf{w}(t+\Delta t)=\bf{w}(t)-\frac{\mathcal{F}^\prime[\bf{w}(t)]}{\mathcal{F}^{\prime\prime}[\bf{w}(t)]}, \label{Newton}
\end{equation}
where ${\bf w}=\{w_1,\dots,w_M\}$ is a $M$-dimensional vector. Here $\mathcal{F}^\prime$ and $\mathcal{F}^{\prime\prime}$ are respectively the first order and the second order gradient of the loss function $\mathcal{F}$. When the loss function reaches the minimum, it should yield $\mathcal{F}=0$, thus the optimization is completed. This process does not require any information of $A_{ij}$ and $|\phi_i\rangle$ as a prior. 

\begin{figure*}
\includegraphics[width=1.0\textwidth]{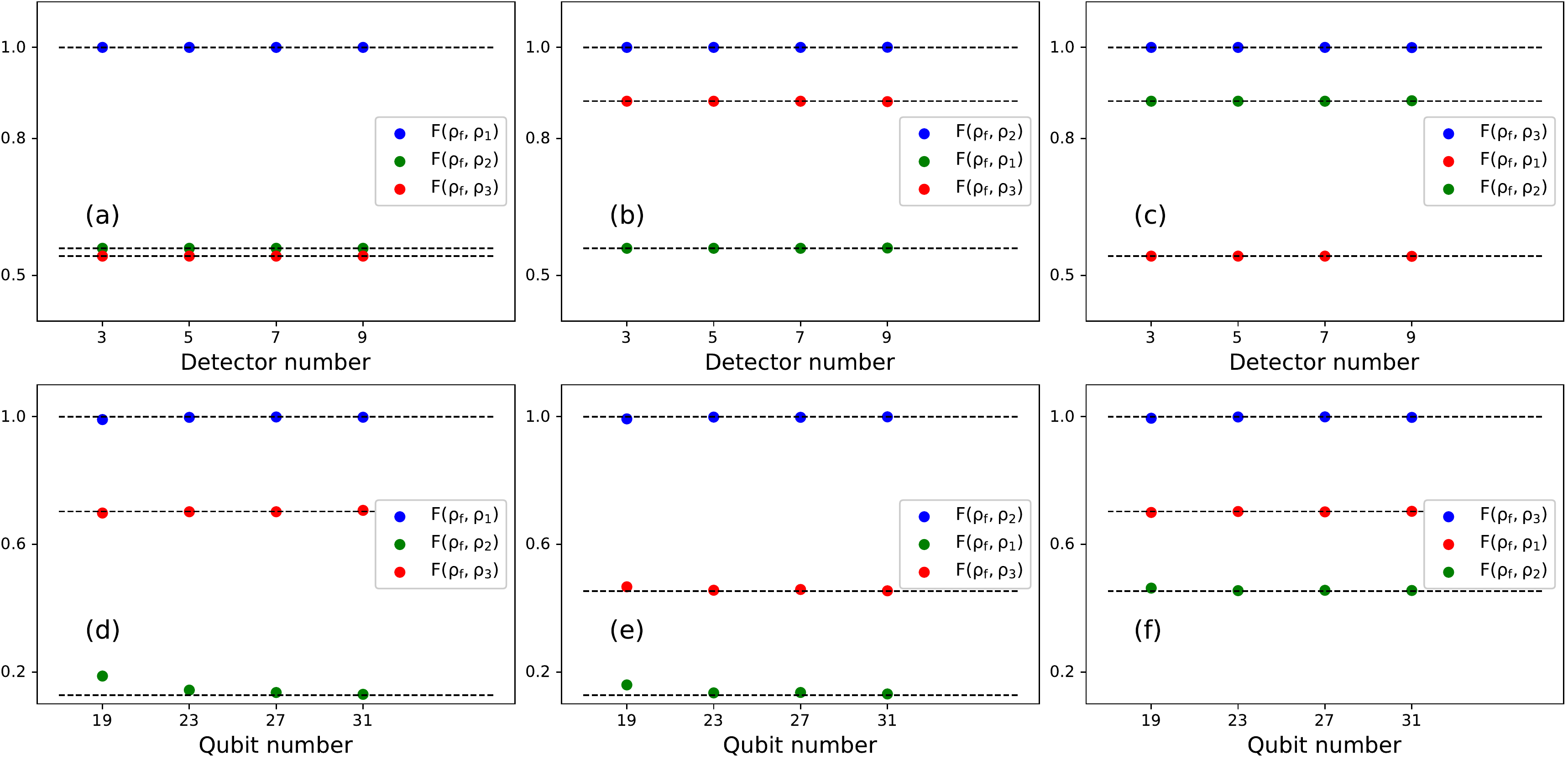}
\caption{(Color online)(a-c) The fidelities versus number of detectors for three different solutions found by the Newton's method. (d-f) The fidelities versus number of spins found by looking for the ground state of the Ising Hamiltonian. Here the number of spins means the number of detectors. In all cases, the solid dots denote the fidelities between the output density matrix $\rho_f$ and three input pure state density matrices, and the dashed lines denote the fidelities between the three input pure state density matrices.   }
\label{results}
\end{figure*}

To test this algorithm, we first randomly generate a set of pure state in the form of
\begin{equation}
|\phi\rangle=\frac{1}{\sqrt{\sum_{k}|c_k|^2}}\sum_{k} c_k|k\rangle,
\label{9}
\end{equation}
where $\{|k\rangle\}$ is a complete set of basis in the Hilbert space $\mathcal{H}_A$ with dimension chosen as $d_A=8$, and $c_k$ is a real coefficient uniformly sampled in the range of $[-5,5]$. To generate $\rho_i$, we randomly sample $A_{ij}$ in the range of $[0,1]$, then normalized under the constraint $\sum_jA_{i,j}=1$. For the example shown in Fig. \ref{results}, we choose three pure states $\rho_{i}=|\phi_i\rangle\langle\phi_i|$ and the fidelities between the three pure states are $F(\rho_1,\rho_2)\doteq 0.56$, $F(\rho_1,\rho_3)\doteq 0.54$ and $F(\rho_2,\rho_3)\doteq 0.88$, where the fidelity is defined as $F(\rho_a,\rho_b)=\text{Tr}\sqrt{\sqrt{\rho_a}\rho_b\sqrt{\rho_a}}$.
 
We then use the Newton's method to solve the q-CPP. $\bf{w}$ is initialized in such a way that $w_i$ ($i=1,\dots,M-1$) are uniformly sampled in the range of $[-2,2]$ and $w_M$ is determined by the constraint $\sum_iw_i=1$. Then, we can reach a convergent solution following Eq. \ref{Newton}. In the example of Fig.\ref{results}, three different $\rho_f$ can be found by the Newton's method depending on different initialization, and their fidelities with $\rho_i$ ($i=1,2,3$) are shown in Fig. \ref{results}(a-c). One can see that there is always one fidelity equalling unity. For instance, for the case Fig. \ref{results}(a), $F(\rho_f,\rho_1)\doteq1$, and $F(\rho_f,\rho_2)$, $F(\rho_f,\rho_3)$ are consistent with $F(\rho_1,\rho_2)$ and $F(\rho_1,\rho_3)$, respectively. This means that the resulting $\rho_f$ recovers $\rho_1$. Similarly, in the cases of Fig. \ref{results}(b) and (c), the resulting $\rho_f$ recovers $\rho_2$ and $\rho_3$, respectively. We have also tried different number of detectors. For the case with three sources, we find that the performance is good as long as the number of detectors is equal or greater than three.

\textit{Mapping to a Hamiltonian Problem.} Since Eq. \ref{loss} is a function of $\{w_j\}$, and if we restrict the value of all $w_j$ to be $\pm 1$, minimizing Eq. \ref{loss} can be regarded as finding the ground state of a Hamiltonian of the Ising spins. If we replace $w_j$ as $\sigma^z_j$, Eq. \ref{loss} can be written into a Hamiltonian form as
\begin{equation}
\hat{H}=\sum\limits_{ijkl}A_{ijkl}\sigma^z_i\sigma^z_j\sigma^z_k\sigma^z_l+\sum\limits_{ijk}B_{ijk}\sigma^z_i\sigma^z_j\sigma^z_k+\sum\limits_{ij}C_{ij}\sigma^z_i\sigma^z_j, \label{H}
\end{equation}
where
\begin{align}
&A_{ijkl}=\text{Tr}(\rho_i\rho_j)(\rho_k\rho_l); \label{A}\\
&B_{ijk}=-\text{Tr}(\rho_i\rho_j)\rho_k-\text{Tr}\rho_i(\rho_j\rho_k);\label{B} \\
&C_{ij}=\text{Tr}\rho_i\rho_j; \label{C}
\end{align}
Here we set the energy unit of the Hamiltonian as unity. In order to satisfy the constraint $\sum_jw_j=1$, we require the number of spin to be odd and the total magnetization to be unity. Note that this Hamiltonian contains four, three and two-body interactions. In this Hamiltonian, the number of sites are equal to the number of detectors. Here it is worth emphasizing that only computing the coefficients listed in Eq. \ref{A}-\ref{C} depends on the pure state Hilbert dimension $d$ of the original quantum problem, and the complicity of the Hamiltonian Eq. \ref{H} itself will not increase as $d$ increases. Given that in the previous discussion of uniqueness of the solution, we prefer to have a large $d$, this is a great advantage of this approach. 

Now the question is whether we can restrict all $w_j$ to be $\pm 1$. Here we make the following statement:

\underline{Statement:} We consider each source ${\bf s}_i$ is a vector, and $M$-number of signal ${\bf x}_j$ ($j=1,\dots,M$) as a mixing of $N$-number of sources ${\bf s}_i$ written as ${\bf x}_j=\sum_{ji}A_{ji}{\bf s}_i$, where all $A_{ji}$ are positive numbers ranging between zero and unity without any other restrictions. We construct ${\bf y}=\sum_{j=1}^{M}w_j{\bf s}_j$ where $w_j$ can only take $\pm 1$. For each given $M$ and for a specified target ${\bf s}_k$, we optimize $\{w_j\}$ ($j=1,\dots,M$) to minimize $|{\bf y}-{\bf s}_k|$ and the minimized value is denoted by $|{\bf y}-{\bf s}_k|^{M}_{\text{min}}$. We state that
\begin{equation}
\lim\limits_{M\rightarrow\infty}|{\bf y}-{\bf s}_k|^{M}_{\text{min}}\rightarrow 0.
\end{equation}  
The meaning of this statement is that, as long as the number of the detector is sufficient, we can always restrict $w_j$ to be $\pm 1$.

\begin{figure}[t]
\includegraphics[width=0.4\textwidth]{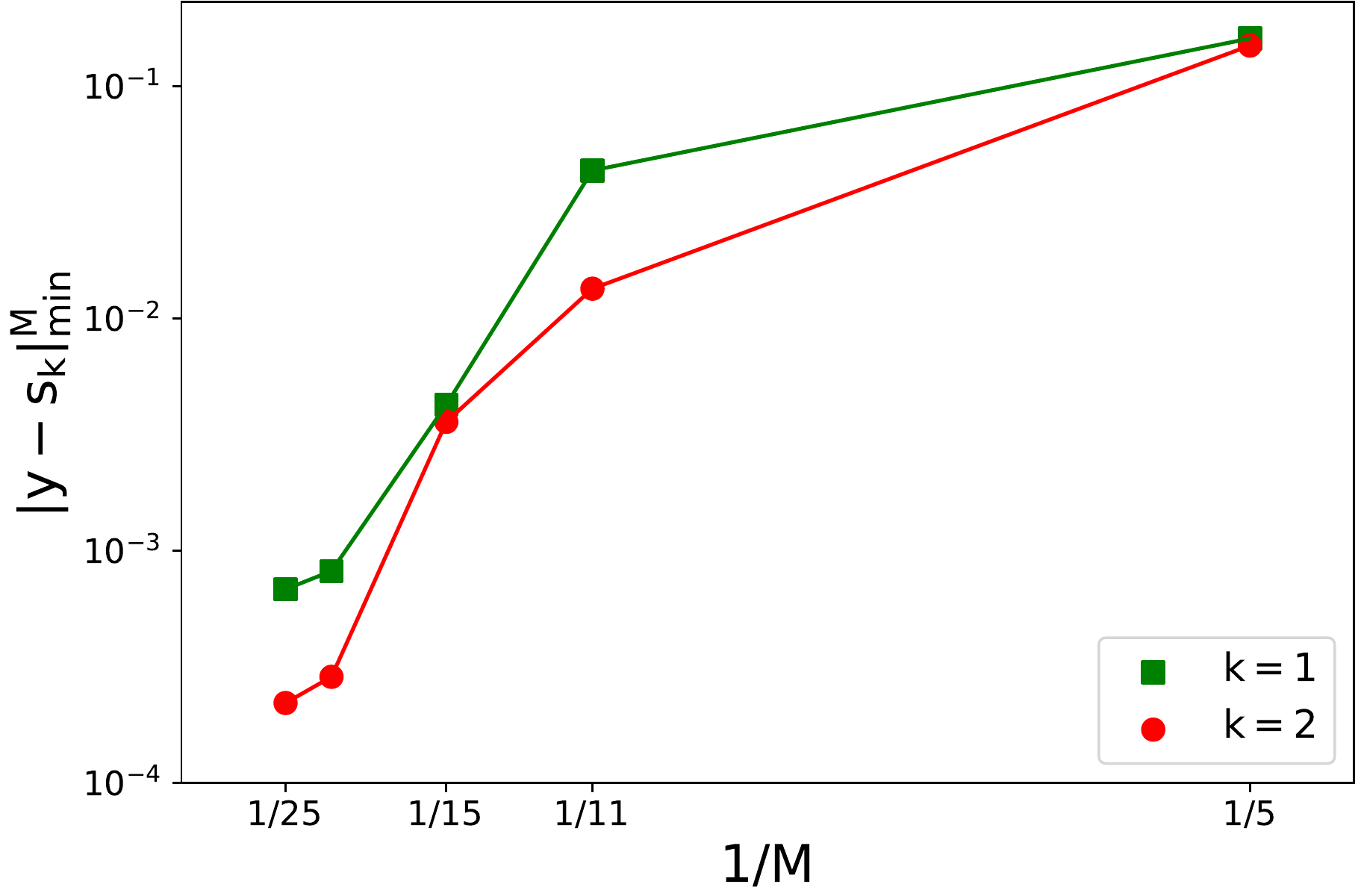}
\caption{Numerical verification of the \textbf{Statement}. Here we consider five sources and $M$ is the number of detectors. }
\label{statement}
\end{figure}

We have verified this statement with numerical simulations. As an example, we consider five sources and each of them is an eight-dimensional vector. As shown in Fig. \ref{statement}, we plot $|{\bf y}-{\bf s}_k|^{M}_{\text{min}}$ as a function of $1/M$ and find that it does converge to zero as $M$ increases.

Now we show the ground state spin configuration of this Hamiltonian can determine the solution of the q-CPP. As $M$ becomes large, the Hilbert space dimension of the Hamiltonian increases and it is hard to solve the ground state by the exact diagnolization. Hence, here we use the classical Monte-Carlo method to find the ground state approximately. In our simulations, the initial temperature is unity which is the same as the energy unit of the Hamiltonian. During the annealing process, the temperature is reduced epoch by epoch, and in each epoch the temperature is reduced by $\frac{1}{n(n+1)}$, where $n$ is the epoch number. In each epoch, we randomly flip the spin for 12000 times with the acceptance probability $P_{\text{accept}}$ given by
\begin{equation}
P_{\text{accept}}=
\left\{\begin{matrix}
1 & E_{t'}<E_t\\
e^{-\beta(E_{t'}-E_t)} & E_{t'}\geq E_t
\end{matrix},\right.
\label{QA_2}
\end{equation}
where $E_{t'}$ and $E_t$ are the eigenenergies after and before randomly flipping the spins, respectively. $\beta=1/(k_\text{b}T)$ is the inverse temperature.
In the simulation, it suffers from the problem of trapped into local minimum. If so, when we regard $\sigma^z_j$ as $w_j$ and reconstruct density matrix $\rho_f=\sum_j w_j\rho_j$, $\rho_f$ may not be a positive definite matrix. Hence, during the annealing process, we simultaneously run two criterions. We require the von Neumann entropy of the reconstructed density matrix, defined as $-\text{Tr}\rho_f\text{log}\rho_f$ to be close to zero, and we also require the reconstructed density matrix to be positive definite. We stop the annealing process at a temperature when these two criterions are well satisfied.

The results of the Monte Carlo calculation are presented in Fig. \ref{results}(d-f). Here we also choose three different input density matrices with fidelity mutually as $F(\rho_1,\rho_2)\doteq 0.13$, $F(\rho_1,\rho_3)\doteq 0.70$ and $F(\rho_2,\rho_3)\doteq 0.45$.
Similar as the results from the Newton's method, depending on different initialization, we can find different spin configurations that can construct three different density matrices $\rho_f$. For instance, for the case shown in Fig. \ref{results}(d), one can find the fidelity $F(\rho_f,\rho_1)$ is very close to unity, and $F(\rho_f,\rho_2)$ and $F(\rho_f,\rho_3)$ are consistent with $F(\rho_1,\rho_2)$ and $F(\rho_1,\rho_3)$, respectively. We can also see that the results get improved as the number of detectors increases. For instance, in Fig. \ref{results}(d), for $M=19$, $23$, $27$, $F(\rho_f,\rho_1)$ are $0.991$, $0.998$ and $0.999$, which gradually approaches unity, and $F(\rho_f,\rho_2)$ are $0.187$, $0.143$ and $0.136$, which gradually approaches the input value $F(\rho_1,\rho_2)$. 

Finally, we should remark that both exact diagnoalization and classical Monte Carlo have limitations for solving the Hamiltonian like Eq. \ref{H}. The most efficient way to find out its ground state is through quantum method, either by quantum simulation or by quantum annealing \cite{annealing}. 

\textbf{Conclusion and Outlook.} 

In summary, we have proposed a quantum analogy of the cocktail party problem. The essential point is to replace the classical signal from each source with quantum data. In the classical problem, it is the independent classical signals that are emitted from different sources, which are mixed and detected by the detectors. In the quantum problem, instead, it is the quantum wave functions that are emitted from different sources, and here the ``independent" means that the wave functions are orthogonal in part of the Hilbert space. The wave functions interfere and the reduced density matrix are observed by detectors. In both cases, the goal is to recover the individual sources from the information of detectors only. 

We also show that solving this problem can be mapped to finding the ground state of an Ising type spin Hamiltonian, and we propose to solve this Hamiltonian by quantum simulator or quantum annealing. We note that, to ensure the uniqueness of the solution, it is preferred to keep both the Hilbert space dimension $d$ of the pure state wave function and the number of detector $M$ large enough. Under this situation, this quantum approach has advantage that on one hand, the complicity of the Hamiltonian does not increase with the increasing of $d$; and on the other hand, the number of Ising spins equal to $M$ and the quantum approach can exhibit its advantage when $M$ is large.   

We envision that one can generalize this quantum version to more complicated situation, such as time-dependent wave function. Similar as that the classical CPP is very useful in classical information processing, we believe the quantum version of CPP can also be useful in quantum information processing.

\textbf{Competing Interests}

The authors declare that there are no competing interests.

\textbf{Author Contribution}

All authors contribute extensively to the entire project. 

\textbf{Data Availability}

Data sets were generated or analyzed during the current study.

\textit{Acknowledgment.} This work is supported MOST under Grant No. 2016YFA0301600 and NSFC Grant No. 11734010.

%\textit{Acknowledgement:}

\end{document}